# Controlled emission of entangled multiphoton states from cascaded quantum wells

Amir Sivan[*] and Meir Orenstein

*Andrew and Erna Viterbi Faculty of Electrical and Computer Engineering, Technion—Israel Institute of Technology, Technion City, Haifa, 3200003, Israel*

We propose a deterministic source of entangled multiphoton states based on spontaneous emission from a ladder of a cascaded quantum well structure. The coupling between the quantum wells enables a many-path evolution with the emission of photon-number combination states in three modes. The tripartite multiphoton state can be used for controlling the entanglement between two multiphoton modes by measuring the third. We further discuss an application as a qubit-pair source with an error-detection ancilla.

Quantum information processing has been at forefront of research for several decades, with much attention given to leveraging the quantum-mechanical properties of collective states of photons, such as squeezing and entanglement, in applications for quantum continuous-variable computation, communication [1-4] and metrology [5-6]. The entanglement of multiphoton systems is a critical resource, but due to the nature of the light matter interactions, the creation of sources of entanglement remains a central challenge. Generation of quantum photonic states is explored in a large variety of systems. The simplest conceptual system is a single two-level system such as an atom or a quantum dot that emits a one-photon Fock state [7]. Using more elaborated energy level schemes in quantum dots, one can also generate pairs of entangled photons [8] and even specific multiphoton states, e.g., a cluster state by repetitive excitations from a quantum dot dark state [9], which requires precise timing of the excitation pulses and is limited by decoherence to entanglement of few photons at cryogenic temperatures. Non-linear optical processes such as spontaneous parametric down conversion can also be utilized to create entangled photon states [10-12]. However, such sources are based on low-probability non-linear interactions and scaling to the many-photon entanglement is problematic. By extending the concept of spontaneous emission from a two-level system, it seems plausible that a deterministic quantum multiphoton state may be realized by an $N$-level system (e.g. $N$ level harmonic oscillator). While deterministic emission of $N$ photons is expected, the photonic quantum state will lack the essential ingredient of entanglement. Even when employing dense indistinguishable collection of two-level systems, which is characterized by highly entangled collective states (Dicke superradiance [13-18]), the $N$-photon emitted state is not reproducing the atomic entanglement [19]. Thus, to generate a genuine spontaneous source for multiphoton states with entanglement we add a crucial ingredient, which is the existence of indistinguishable multiple emission paths in the ladder system, that are missing in the above examples. We thus devise here a realizable energy-ladder based system which supports multiple possible spontaneous photon emission sequences as a source for multiphoton entanglement between the emitted photon modes.

The source is based on the spontaneous emission of quantum multiphoton states (MPS) from cascaded coupled quantum wells (CQW). We show that the spontaneous process results in highly entangled three modes MPS and furthermore that a two-mode entanglement can be switched by a post-selection measurement. We demonstrate a possible application as a source of qubit pairs with an error detection ancilla.

The CQW structure consists of a repetitive structure of identical quantum wells (QWs) and barriers with energies that are relatively biased such that the ground state of each QW is aligned and coupled with the excited state of the subsequent one as described in Fig. 1a. After an initialization by excitation of an electron to the top QW, the system evolves down the energy ladder of the cascade through spontaneous emission of photons and inter-well electron tunnelling. This process is reminiscent of quantum-cascade lasing [20], with the emphasis that here the electron tunneling between adjacent QWs is coherent (rather than, e.g., phonon-assisted). We remark that tailoring and realizing such a structure by semiconductor epitaxy is well established and the emission process is not limited to cryogenic temperatures and therefore can be potentially used in many practical scenarios.

The entanglement between the MPS occurs between the three emitted photonic modes (frequencies), that originate from the many-path evolution in the CQW energy ladder. Each energy level of the original QWs is split to higher and lower sub-levels due to the inter-well coupling. The splitting, which depends on structure parameters, defines four possible transitions within each QW that correspond to emissions in three distinct frequencies (Fig. 1b) and result in MPS in three inter-entangled modes. Furthermore – this tripartite entangled state provides us with a control knob. When the photon number of the central mode is measured after the CQW has reached its ground state, the two remaining modes MPS are projected onto either as an entangled or a separable state, depending on the parity of the measured photon number. Thus, we get the desirable quantum source which is both deterministic (the CQW system signal us when it reaches the ground state and emits a deterministic number of photons) and controlled by "enabling" or "disabling" multiphoton entanglement by post-selection.

---
[*] amirsi@campus.technion.ac.il

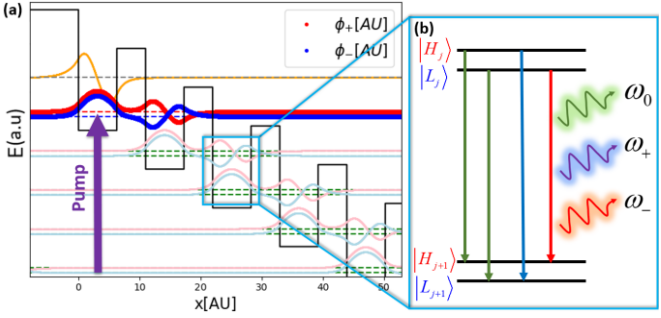

Fig. 1. (a) Schematics of the cascaded quantum well structure and the supported electronic states. Each quantum well (QW) supports ground and excited levels, and each level is split into two sublevels resulting from coupling to the neighboring QW. The exciting pump to the first QW (purple arrow) can be either optical or electronic. (b) QW sublevel schematic, with the four possible interlevel transitions emitting photons in three possible modes, $\omega_0$, $\omega_-$ and $\omega_+$.

*QW coupling scheme.* In our cascade structure all the separated QWs are identical and each supports two energy levels (Fig. 1a). To facilitate the tunnelling between adjacent QWs we shift the potential well of each consecutive QW by the level energy spacing, such that the ground energy level of a QW is aligned with the excited energy of the following QW. In practice, this can be achieved by proper material compositions or by applying an electric field to the CQW structure. Starting with the single asymmetrical QWs we solve the excited and ground state wavefunctions $\phi_1$ and $\phi_0$, respectively (see Supplementary Material I [21]). Then, we employ the coupled-mode theory (CMT) to calculate the superposition wavefunctions $\phi_\pm = a_\pm \phi_1 + b_\pm \phi_0$ that result from the inter-well coupling [22]. Owing to this coupling, each original energy level is split by $\delta E$. When considering cascaded identical QWs, the four energy levels (two excited and two ground levels) define four radiative transitions in the $j$'th QW, namely $|H_j\rangle \to |H_{j+1}\rangle$, $|L_j\rangle \to |L_{j+1}\rangle$, $|L_j\rangle \to |H_{j+1}\rangle$ and $|H_j\rangle \to |L_{j+1}\rangle$ where the state $|H_j\rangle$ ($|L_j\rangle$) denotes the higher (lower) sublevel of the ground state of the $j$'th QW which is also describes the higher (lower) sublevel of the excited state of the $(j+1)$'th QW. The first two transitions from the list above correspond to a photon with angular frequency $\omega_0 = \Delta E / \hbar$, whereas the other two respective transitions to angular frequencies $\omega_+ = (\Delta E + \delta E)/\hbar$ and $\omega_- = (\Delta E - \delta E)/\hbar$ (Fig. 1b). The transition probabilities and spontaneous emission rates are derived from the dipole transition moment by Fermi's golden rule.

*State vectors and photon emission evolution from the CQW.* After each photon emission, the state vector composed of the CQW and the photonic state evolves. The initial excitation of the system is to the ground state of the first QW with population amplitudes $c_H, c_L$ of the high and low sublevels, respectively. The respective state vector is $|\psi_1\rangle = (c_H|H_1\rangle + c_L|L_1\rangle) \otimes |0\rangle_- |0\rangle_0 |0\rangle_+$. Here we defined three photon states – one for each of the three possible emission modes, where vector $|0\rangle_x$ defines the vacuum in the mode corresponding to $\omega_x$. For brevity we denote the photon state $|l,m,n\rangle \equiv |l\rangle_- |m\rangle_0 |n\rangle_+$ for photon numbers $(l,m,n)$ in the respective modes. It is important to note that we separate between the "interaction zone", where the CQW system evolves and releases photons, and the "far field zone" where we measure or use the MPS. In the interaction zone we employ the Markovian (bad cavity) approximation, i.e., that the photon residence time within the structure is negligible, so that the emitted photons do not act back on the CQW. This means that in this zone, the photonic state is reset to vacuum after each emission. In the far field zone, we effectively store the arriving photons in three photon cavities, each for each photon mode. We use or measure these three multiphoton modes after the cascade is terminated.

After the first photon is emitted, the state vector evolves to

$$|\psi_1\rangle \to |\psi_2\rangle = |H_2\rangle(c_H c_{++}|0,1,0\rangle + c_L c_{+-}|0,0,1\rangle)$$
$$+ |L_2\rangle(c_H c_{-+}|1,0,0\rangle + c_L c_{--}|0,1,0\rangle)$$

where the amplitudes $c_{++}, c_{--}, c_{+-}$ and $c_{-+}$ are derived from the transition probability amplitudes. More generally, immediately after each photon emission, we evolve the states that denote the sublevels of the ground state in the $j$'th well $|H_j\rangle$ and $|L_j\rangle$ as

$$|H_j\rangle|l,m,n\rangle \to \frac{1}{\sqrt{c_{++}^2 + c_{+-}^2}} \times$$
$$(c_{++}|H_{j+1}\rangle \hat{a}^\dagger_{\omega_0} + c_{+-}|L_{j+1}\rangle \hat{a}^\dagger_{\omega_+})|l,m,n\rangle$$
$$|L_j\rangle|l,m,n\rangle \to \frac{1}{\sqrt{c_{-+}^2 + c_{--}^2}} \times \quad (1)$$
$$(c_{-+}|H_{j+1}\rangle \hat{a}^\dagger_{\omega_-} + c_{--}|L_{j+1}\rangle \hat{a}^\dagger_{\omega_0})|l,m,n\rangle$$

for $l+m+n = j$ and $j \leq N$, where $\hat{a}^\dagger_x$ is the raising operator of a photon in the far field mode $\omega_x$ defined as $\hat{a}^\dagger_x |p\rangle_x = |p+1\rangle_x$ for a photon number $p$.

Describing the evolution by updating the state vector after every emission has the advantage of simplicity while maintaining all the information regarding the different evolution paths at each photon emission event, which is convenient for describing systems with complicated dynamics [23,24]. By repeatedly applying the state evolution mapping, one obtains the general state vector after all photons were emitted. In the last transition, where both the $|H_{N-1}\rangle$ and $|L_{N-1}\rangle$ sublevels decay to the final common ground state of the last CQW, the transition frequencies are slightly different from the $\omega_+$ and $\omega_-$ modes, but we still accumulate these photons in the $\omega_+ > \omega_0$ and $\omega_- < \omega_0$ effective cavities, respectively. The multi-path evolution is described graphically in Fig. 2, which represents a nonreciprocal hopping scheme in a tight-binding-like crystal. After the CQW system signals us that it reached its final ground state, we know that $N$ photons were emitted from the CQW (we assume low probability of nonradiative processes).

From (1) it can be seen that any phase accumulated by a state $|H_j\rangle$ or $|L_j\rangle$ will become a global phase of the superposition $j+1$ states after the $j$'th photon was emitted. Consequently, at the end of the CQW process the system will have accumulated the same global phase regardless of which quantum path was taken. Because of this, if the PND measures the photon numbers $(l,m,n)$ corresponding to the $\omega_-, \omega_0$ and $\omega_+$ modes, respectively, there would be no physical measurement that would differentiate between the different possible paths that produce the $(l,m,n)$ photon numbers.

The probability that the three photon numbers are $(l,m,n)$ after $N$ transitions have occurred, can be calculated from (1). We will denote the probability to measure the photon number combination $(l,m,n)$ as $f(l,m,n)$. It is easy to verify that these fulfill the summation identity $\sum_{\{l,m,n\}\in K} |f(l,m,n)| = 1$ where $\{K : \forall l,m,n : l+m+n = N, |l-n| \geq 1\}$. These probabilities describe the $N$-step distance of a three-dimensional (3D) random walk with drift under constraints on the $l, n$ dimensions, which are derived from the transition rules between high (low) and low (high) energy sublevels in adjacent QWs, wherein each step describes a photon emission with a mode dictated by the direction of the step.

*Entanglement of the multi-photonic state.* The many-path dynamics of the CQW results in superpositions of MPS, each consists of the weighted photon number states of the three modes. When the process has reached the CQW ground level, the final photonic state of the system in the photon number description is given by the superposition

$$|\psi_N\rangle = \sum_{\{l,m,n\}\in K} \sqrt{f(l,m,n)} |l,m,n\rangle \quad (2)$$

Considering the density matrix $\rho_N = |\psi_N\rangle\langle\psi_N|$, from (2) and from the definition of $f(l,m,n)$ it is clear that $Tr\rho_N = 1$ and $rank\rho_N = 1$, and it can be readily proven that $\rho_N^2 = \rho_N$ and therefore $\rho_N$ represents a pure state that is non-separable as is evident from (1).

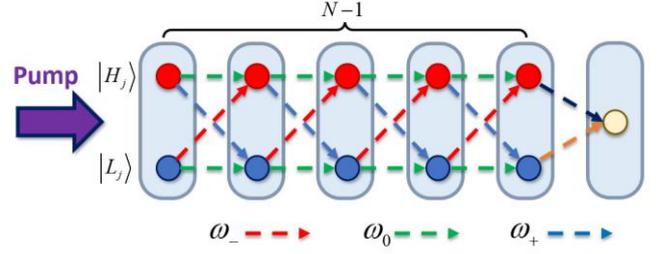

Fig. 2. Schematic of the cascaded quantum well structure evolution as hopping in a tight-binding model. Each unit cell describes a quantum well (QW) in the cascade. The red (blue) circle in the *j*'th QW denotes the respective $|H_j\rangle$ ($|L_j\rangle$) sublevel. The system is initialized by a pump to the leftmost unit cell. The dashed arrows denote the transitions between the sublevels associated with the modes that color coded red, green and blue for $\omega_-, \omega_0$ and $\omega_+$ respectively. The last QW contains the ground level of the CQW, and all paths converge to it via either the indigo or orange transitions.

This state can be described as a discrete finite distribution of number states with three parameters – the three photon modes. In Fig. 3, we demonstrate the mutual probability distributions (MPD) of the MPS of $\omega_+, \omega_-$ given that the CQW system has reached its final ground state.

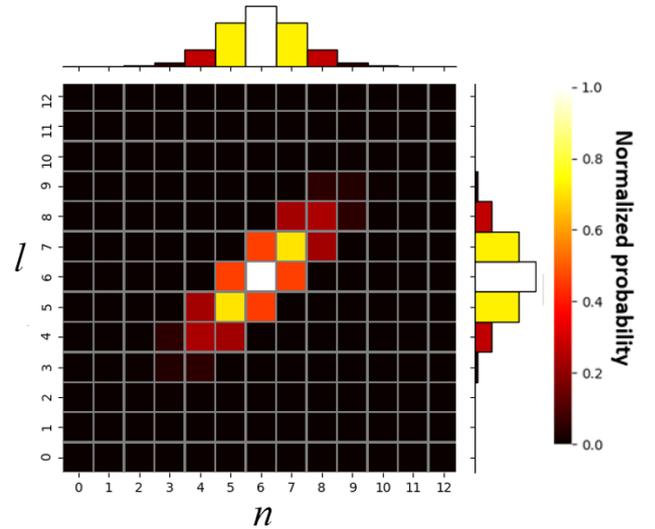

Fig. 3. Mutual probability distribution of the photon numbers of the $\omega_+$ and $\omega_-$ modes photons, in a structure comprised of $N = 22$ cascaded quantum wells. The histograms describe the marginal probabilities of the photon numbers of the respective MPS.

The main result of this letter is the connection between the three mode MPS in the final state. From Fig. 3, we clearly see that the condition on $l$ and $n$ constraints the MPD such that after a measurement of the photon number of any single-mode MPS the photon numbers of the remaining two single-mode

MPS are completely determined. A particularly interesting case is when the photon number state of mode $\omega_0$ is measured. The remaining combined two-mode MPS with (the unmeasured) numbers $l$ and $n$ is conditioned by the parity of $s = N - m$,

$$\langle m | \psi_N \rangle = |k,k\rangle, \quad s \bmod 2 = 0 \quad (3)$$

$$\langle m | \psi_N \rangle = \alpha_m |k,k+1\rangle + \beta_m |k+1,k\rangle, \quad s \bmod 2 = 1 \quad (4)$$

with $k = \lfloor s/2 \rfloor$, where the vectors in the RHS are denoted as $|l,n\rangle$. The coefficients $\alpha_m, \beta_m \in \mathbb{R}$ are given by (2) for the appropriate $m$.

The state $\langle m | \psi_N \rangle$ is a bipartite pure state and thus its non-separability is a necessary and sufficient condition for its entanglement. Since the total emitted photon number $N$ is given, equations (3)-(4) reveal the interesting property that the measurement result of $m$ dictates not only the photon numbers in the two unmeasured MPS, but also whether they are separable (3) or entangled (4) Fock states. In the latter case, measurement of the photon number in one of these two modes would collapse the state in the other mode to the "complementary" Fock state.

*Application example: a qubit source with ancillary error-detection.* For elucidating a simple application of this source, we describe a setting where two multiphoton modes from the CQW system are used as a pair of qubits and the third mode as an error detection ancilla. This enables a non-destructive measurement, which is highly desired for quantum computation or communications. Specifically, the measurement of the parity of the photon number in $\omega_0$ mode, yields a logical XOR operation between the parities of the two remaining multiphoton modes. Each of the two qubits is encoded in its respective mode ($\omega_-$ or $\omega_+$) by either an even or odd photon number and the corresponding logical qubit basis is the parity of this photon number. For even $N$, the logical value (parity) measured for the $\omega_0$ mode will correspond to the logical result of a XOR operation acting on the logical values of the two qubits at the $\omega_-$ and $\omega_+$ modes. In this basis, a parity measurement resulting in $|0\rangle_m$ yields a superposition of $|0,0\rangle_{l,n}$ and $|1,1\rangle_{l,n}$, whereas $|1\rangle_m$ results in a superposition of $|0,1\rangle_{l,n}$ and $|1,0\rangle_{l,n}$. For an odd number of CQWs, the respective logical $m$-qubits are flipped and the result is of a logical NXOR. This effectively enables a qubit parity test of the $(l,n)$ pair by measurement of the ancilla qubit $m$.

*Conclusion.* We propose a source of entangled multiphoton states. The entanglement is generated via a multipath spontaneous emission process in a cascaded quantum well structure. It is important to emphasize that such source of entangled photons can be realized by using existing semiconductor materials, epitaxial growth and fabrication technologies. The typical tunneling rates in the regime of picoseconds and light emission rates in the nanoseconds are well within the approximation that we applied in our model. In our analysis we ignored the nonradiative recombination processes – which may deteriorate the multiphoton state. However, the management of the nonradiative processes is well established and their relative reduction can be achieved by shorter wavelength emission, lower temperatures, or by enhancement of the emission rate – e.g., by plasmonic structures. Internal quantum efficiency of radiative processes in LEDs close to 100% were demonstrated [25].

The CQW has several interesting properties beyond emission of multiphoton entangled states. Upon post-measurement of the central mode photon number, the two other modes are projected to either an entangled or an unentangled photon number states depending on the parity of the measured photon number.

We demonstrated that in a photon number parity logical qubit base, the photon numbers of the three MPS are interrelated as the inputs and output of a XOR (or NXOR) gate. This enables the proposed CQW structure to function as a source of multiphoton qubit pairs with an error detection ancilla, which is an important resource for quantum communication and computation applications.

The proposal presented in this letter can be further expanded by introducing QWs with more energy levels or different oscillator strengths, as well as more intricate inter-QW coupling schemes. Such design modifications will result in photonic states with even richer dynamics and entanglement properties.

1. J. W. Pan, Z. B. Chen, C. Y. Lu, H. Weinfurter, A. Zeilinger, and M. Żukowski, Multiphoton entanglement and interferometry. Reviews of Modern Physics, 84(2), 777 (2012).
2. U. L. Andersen, J. S. Neergaard-Nielsen, P. Van Loock, and A. Furusawa, Hybrid discrete-and continuous-variable quantum information. Nature Physics, 11(9), 713-719 (2015).
3. A. I. Lvovsky and M. G. Raymer, Continuous-variable optical quantum-state tomography. Reviews of modern physics, 81(1), 299 (2009).
4. S. L. Braunstein and P. Van Loock, Quantum information with continuous variables. Reviews of modern physics, 77(2), 513 (2005).
5. V. Giovannetti, S. Lloyd, and L. Maccone, Advances in quantum metrology. Nature photonics, 5(4), 222-229 (2011).
6. S. Pirandola, B. R. Bardhan, T. Gehring, C. Weedbrook, and S. Lloyd, Advances in photonic quantum sensing. Nature Photonics, 12(12), 724-733 (2018).

7. P. Michler, A. Kiraz, C. Becher, W. V. Schoenfeld, P. M. Petroff, L. Zhang, and A. Imamoglu, A quantum dot single-photon turnstile device. science, 290(5500), 2282-2285 (2000).
8. N. Akopian, N. H. Lindner, E. Poem, Y. Berlatzky, J. Avron, D. Gershoni, and P. M. Petroff, Entangled photon pairs from semiconductor quantum dots. Physical review letters, 96(13), 130501 (2006).
9. I. Schwartz, D. Cogan, E. R. Schmidgall, Y. Don, L. Gantz, O. Kenneth, and D. Gershoni, Deterministic generation of a cluster state of entangled photons. Science, 354(6311), 434-437 (2016).
10. H. Hübel, D. R. Hamel, A. Fedrizzi, S. Ramelow, K. J. Resch, and T. Jennewein, Direct generation of photon triplets using cascaded photon-pair sources. Nature, 466(7306), 601-603 (2010).
11. L. Caspani, C. Xiong, B. J. Eggleton, D. Bajoni, M. Liscidini, M. Galli, and D. J. ...Moss, Integrated sources of photon quantum states based on nonlinear optics. Light: Science & Applications, 6(11), e17100-e17100 (2017).
12. C. Zhang, Y. F. Huang, B. H. Liu, C. F. Li, and G. C. Guo, Spontaneous parametric down-conversion sources for multiphoton experiments. Advanced Quantum Technologies, 4(5), 2000132 (2021).
13. R. H. Dicke, Coherence in spontaneous radiation processes, Phys. Rev. 93, 99 (1954).
14. R. Bonifacio, P. Schwendimann, and F. Haake, Quantum statistical theory of superradiance. I. Physical Review A, 4(1), 302 (1971).
15. M. Gross and S. Haroche, Superradiance: An essay on the theory of collective spontaneous emission. Physics reports, 93(5), 301-396 (1982).
16. A. Sivan and M. Orenstein, Enhanced superradiance of quantum sources near nanoscaled media. Physical Review B, 99(11), 115436 (2019).
17. A. Asenjo-Garcia, M. Moreno-Cardoner, A. Albrecht, H. J. Kimble, and D. E. Chang, Exponential improvement in photon storage fidelities using subradiance and "selective radiance" in atomic arrays. Physical Review X, 7(3), 031024 (2017).
18. S. John and T. Quang, Localization of superradiance near a photonic band gap. Physical review letters, 74(17), 3419 (1995).
19. E. Wolfe and S. F. Yelin, Certifying separability in symmetric mixed states of n qubits, and superradiance. Physical review letters, 112(14), 140402 (2014).
20. J. Faist, F. Capasso, D. L. Sivco, C. Sirtori, A. L. Hutchinson, and A. Y. Cho, Quantum cascade laser. Science, 264(5158), 553-556 (1994).
21. See Supplementary material at XXX.
22. S. L. Chuang and B. Do, Electron states in two coupled quantum wells—A strong coupling-of-modes approach. Journal of applied physics, 62(4), 1290-1297 (1987).
23. C. Cohen-Tannoudji and S. Reynaud, Dressed-atom description of resonance fluorescence and absorption spectra of a multi-level atom in an intense laser beam. Journal of Physics B: Atomic and Molecular Physics (1968-1987), 10(3), 345 (1977).
24. V. Bužek and P. L. Knight, *I: Quantum interference, superposition states of light, and nonclassical effects*. (Progress in optics, Elsevier, 1995), Vol. 34, p. 1-158.
25. I. Schnitzer, E. Yablonovitch, C. Caneau, and T. J. Gmitter, Ultrahigh spontaneous emission quantum efficiency, 99.7% internally and 72% externally, from AlGaAs/GaAs/AlGaAs double heterostructures. Applied physics letters, 62(2), 131-133 (1993).

# Controlled emission of entangled multiphoton states from cascaded quantum wells

Amir Sivan[*] and Meir Orenstein

*Andrew and Erna Viterbi Faculty of Electrical and Computer Engineering, Technion—Israel Institute of Technology, Technion City, Haifa, 3200003, Israel*

## I. QW COUPLING SCHEME

In this section we will provide details on the design of the identical coupled QWs under the requirements detailed in the main text, namely that exactly two levels (ground and excited) are supported in each QW, and that the ground level energy of the j'th QW is equal the excited level energy of the (j+1)'th QW. We consider a cascade of asymmetric QWs such that the right barrier potential is biased by $-b$ ($b \in \mathbb{R} \geq 0$) with respect to the left barrier (see Fig. 1a in the main text). As mentioned in the main text, these requirements impose a relationship between the QW width $D$ and the bias $b$. We employ a formalism similar to that of [1]. The eigenenergies $E_n$ for each QW are obtained numerically from the transcendental equation

$$\kappa D = \tan^{-1}(\nu/\kappa) + \tan^{-1}(\delta/\kappa) + n\pi \tag{S1}$$

with

$$\nu = \sqrt{\frac{2m^*}{\hbar^2}(V_1 - E)}, \quad \kappa = \sqrt{\frac{2m^*}{\hbar^2}(V_2 - E)}, \quad \delta = \sqrt{\frac{2m^*}{\hbar^2}(V_1 - b - E)} \tag{S2}$$

for $n \geq 0$, with $n = 0$ denoting the ground level energy. Here $\nu, \kappa, \delta$ are the constants for the left barrier, quantum well, and right barrier. $V_1$ is the potential of the left barrier and $V_2$ is the potential of the well. We will for brevity $2m^*/\hbar^2 = 1$ in the appropriate units. The adjacent QW from the right is identical up to a spatial translation $L$ along the $x$ axis, and a potential offset such that $V_1 \to V_1 - b$ and $V_2 \to V_2 - b$ as described in Fig. It can be deduced from the above that the eigenenergies of the second QW satisfy $E'_n = E_n - b$. As mentioned, we require that the transcendental equation will not have solutions for $n > 1$ and that $E'_1 = E_0$. These may be restated as $\Delta E \equiv E_1 - E_0 = b$.

To find the wavefunctions of the cascaded structure, we use CMT, following the notations and procedure developed in Ref. [1]. We assume that because all QWs support exactly two states, each eigenstate of the entire system is comprised of a superposition of only two states $\phi_{j,0}, \phi_{j+1,1}$ – the ground state of the j'th QW and the excited state of the (j+1)'th QW. The two states supported by the coupled j'th and (j+1)'th QW pair are therefore given by

$$\phi_{j+1,\pm} = a_\pm \phi_{j,0} + b_\pm \phi_{j+1,1} \tag{S3}$$

With $a_\pm, b_\pm$ obtained from the CMT. Each (j,j+1) pair of QWs shares two such states centered around the ground level energy of the j'th QW, as portrayed in Fig. 1a in the main text.

References

1. S. L. Chuang and B. Do, Electron states in two coupled quantum wells—A strong coupling-of-modes approach. Journal of applied physics, 62(4), 1290-1297 (1987).

---

[*] amirsi@campus.technion.ac.il